\documentstyle[prc,aps,preprint]{revtex}
\author{B. J. Cole}
\address{Department of Physics, University of the Witwatersrand,
     WITS 2050, South Africa}
\author{H. G. Miller and R. M. Quick}
\address{Department of Physics,
        University of Pretoria, Pretoria 0002, South Africa}

\title{Comment on ``Specific Heat and Shape Transitions in Light {\it
sd}
Nuclei''}
\date{\today
}
\draft
\begin{document}

\maketitle
\begin{abstract}
This comment re-examines the origin of structure seen in the
computed specific heat of finite nuclei.  In a recent paper,
Civitarese and Schvellinger suggest that such structure is due
to model-space truncation in the calculations.  We reaffirm our
conclusion that the structure is caused by a
collective-to-non-collective
phase transformation at low temperatures,
signaled by a change in the nuclear level density below 10 MeV
excitation energy.
\end{abstract}


In a recent paper Civitarese and Schvellinger~\cite{CS94} have
presented the
results of calculations of the thermal properties of eleven even-even
{\em sd%
} nuclei in the SU3 model of Elliott~\cite{E58}. Within the framework
of
this model they calculated the specific heat and the ensemble average
of the
intrinsic quadrupole moment of these nuclei  in the canonical
ensemble.
Since the latter quantity does not vanish as the temperature is
raised,
although a peak is observed in the specific heat, they claim to have
presented evidence against the occurrence of phase transitions in
light
nuclei. Similar results have also been obtained by Dukelsky $et.\
al.$~\cite
{DPR91}.  Moreover, Civitarese and Schvellinger suggest that their
results
are in contradiction to those of our earlier work~\cite{MCQ89}. This
is not
the case, and in the present comment we simply wish to set the record
straight
with regard to our arguments in favor of the existence of a low
temperature
collective-to-non-collective phase transition in nuclei~\cite
{MCQ89,QDCM91,RQM92,CRM92,YMQ93}.

In a series of papers we have attempted to compare finite-temperature
Hartree-Fock (FTHF), or approximate canonical calculations, with
canonical
calculations which use the exact eigenstates of the same realistic
effective
Hamiltonian operator~\cite{VY77,SMBV78} in the same model space~\cite
{MCQ89,QDCM91,RQM92,CRM92,QMC89,CQM89,MQC89,QCM92}. For this reason
we have
restricted our calculations to the $sd$ shell, where exact
shell-model
calculations are possible. In the FTHF calculations in both
$^{20}$Ne~\cite
{RQM92} and $^{24}$Mg~\cite{MQC89} the ensemble average of the
quadrupole
moment vanishes, and this was taken to indicate the occurrence of a
shape
transition~\cite{MZP74,MQBV86}. However, as we have previously
pointed out ~%
\cite{MCQ89,QDCM91,RQM92,MQC89}, this behavior appears to an artifact
of the
finite-temperature mean-field calculation and depends on the volume
of the
system~\cite{YM92,YM94}. The canonical-ensemble average of the
quadrupole
moment~\cite{MQC89} or the Hill-Wheeler deformation parameter $\beta
$~\cite
{RQM92} calculated with exact shell model eigenstates does not
vanish, in
agreement with results obtained in the SU3 model~\cite{CS94,DPR91}.
More
recent calculations of the ensemble average of the quadrupole moment
squared $%
Q^{[2]}\cdot $ $Q^{[2]}$ indicate that it is discontinuous in the
FTHF
approximation, while no discontinuity is observed in the canonical
calculations~\cite{CMQ94}. Here it should also be noted that for
light nuclei
the thermal fluctuations in the quadrupole moment and $\beta $ are
large and,
when taken into account, wash out any differences between the
predictions of the
finite-temperature mean-field calculations and the exact canonical
calculations~\cite{MQC89,LA84}. For this reason we have expressed
some
concern about attempts to confirm experimentally the existence of
such shape
transitions~\cite{YMQ93}.

The question now arises: Does a low-temperature phase transition (or,
more
correctly, a phase transformation) take place in nuclei? Clearly it
cannot be a
true phase transition because these take place only in infinite
systems.
However, if we use as operational definition of a phase transition in
a
finite system, the remnant of the true phase transition which would
occur in
the thermodynamic limit of the finite system~\cite{FGN79,RP85}, then
we feel
that there is evidence for a non-collective-to-collective phase
transition in
finite nuclei. One of the universal features of finite nuclei is an
abrupt
change in the density of states at excitation energies of 10 MeV or
less
\cite{GC65}. At lower excitation energies the spectrum of most nuclei
is
sparse and dominated by a relatively small number of collective
states. With
increasing excitation energy, the independent particle degrees of
freedom
dominate and the density of states grows exponentially. Moreover,
since the nuclear
force is short-ranged and saturates rather quickly one would expect
that such
a change in the many-body level density might also occur in nuclear
matter. A
recent semi-empirical determination of the properties of nuclear
matter~\cite
{DHMMT93} has indeed provided evidence for such a low-temperature
phase
transition, which is consistent with the predictions of realistic
finite-temperature BCS calculations~\cite{BCLL90} .

We have therefore suggested that the peak seen in the specific heat,
$C$, in
both  FTHF calculations and  canonical calculations for finite
nuclei~\cite
{MCQ89,QDCM91} is indicative of this collective-to-non-collective
phase
transition. In both cases, it is useful to note~\cite{P84,G90} that
the
specific heat is given by
\begin{eqnarray*}
C &=& \frac{\partial E}{\partial T} \\
  &=& T\frac{\partial S}{\partial T}
\end{eqnarray*}
where $E$ is the ensemble average of the energy and $T$ is the
temperature.
The entropy is given by
$$
S=\log \rho
$$
where $\rho $ is the many-body density of states. Peaked structures
in the
specific heat, which we have identified as signatures of the
aforementioned
phase transition, can be directly related to abrupt changes in the
many-body
level density~\cite{MCQ89,QDCM91}, which are clearly due to dynamical
effects~\cite{CRM92}. Clearly one must be careful to ascertain that
any
peaked structure in $C$ does not arise from the finite size of the
model
space~\cite{DPR91,CDZ89}. For this reason we have performed canonical
calculations of $C$ using not only the eigenstates of an $sd$ shell
effective interaction, but also the experimental nuclear structure
information~\cite{MCQ89}. In $^{20}$Ne in both cases a peaked
structure is
observed in $C$, which agrees with the FTHF predictions~\cite{MCQ89}.
If the
continuum contribution is taken account~\cite{MCQ89,CRM92,CMQ90}, the
peak
in $C$ remains, although the behavior of $C$ changes at higher
temperatures.
Furthermore, we have also shown~\cite{QDCM91} that the FTHF does very
good
job of approximating the change in the many-body level density seen
in the
shell-model calculations. This change in the spectral distribution,
or
normalized level density, is also seen in the SU3 model calculations
in
Figure 2 of Ref~\cite{CS94}.

In conclusion, we would like to point out that the results in
Ref~\cite{CS94}%
, in our opinion, do not disagree with our previous work but, on the
contrary, seem to support our conclusions with respect to the
existence of a
low-temperature collective-to-non-collective phase transition in
nuclei.

\acknowledgements
The support of the Foundation for Research Development of South
Africa is
gratefully acknowledged.


\end{document}